\documentclass[10pt,oneside]{article}
\usepackage{latexsym}
\usepackage{graphics}
\usepackage{epsf}
\usepackage[inline]{asymptote}
\usepackage{amsmath}
\usepackage{amssymb}
\usepackage{amsfonts}
\usepackage[T1]{fontenc}
\usepackage{relsize,exscale}
\usepackage{ntheorem}
\theoremseparator{:}
\renewcommand{\Bbb}{\mathbb}
\newcommand{\RR}{\Bbb{R}}
\newcommand{\NN}{\Bbb{N}}
\newcommand{\ZZ}{\Bbb{Z}}
\newcommand{\QQ}{\Bbb{Q}}

\newtheorem{lemme}{Lemme}[section]

\def\supp\mathop{supp}
\evensidemargin=\oddsidemargin
\begin{document}
\title{A proof of the convexity of the set of lamination parameters}
\author{Jean-Luc AKIAN}

\maketitle
\begin{center}

ONERA, Universit\'{e} Paris Saclay F-92322 Ch\^{a}tillon, France

Tel: +33-1-46-73-46-41; Fax: +33-1-46-73-41-43

E-mail address: jean-luc.akian@onera.fr

\end{center}

\maketitle
\begin{abstract}
In this paper we show that the proof of the convexity of the set of lamination parameters given by J.L. Grenestedt and P. Gudmundson, which is extensively cited in the literature, is not correct. We give a proof of the convexity of this set when the class of functions which gives the layup angle as a function of the through-the-thickness coordinate is the set of step functions.
\end{abstract}

{\bf keywords}: lamination parameters, convexity
\section{Introduction}
Lamination parameters are extensively used for the layup optimization of laminated composite structures instead of the ply thicknesses and the layup angles. These parameters are integrals through the thickness of functions of the layup angles of the different plies of the laminate and their number is small (12). A key result for layup optimization is the convexity of the set of lamination parameters. A remark is that when one speak of the convexity of the set of lamination parameters, one must specify the class of functions which gives the layup angle as a function of the through-the-thickness coordinate. For a composite laminate this function should be a step function (that is to say a piecewise constant function). The paper \cite{Grenestedt} gives a proof of the the convexity of the set of lamination parameters and is extensively cited in the literature for this proof. But the proof of this paper is not correct as will be shown in the present paper.
In the present paper we give a proof of the convexity of the set of lamination parameters when the class of functions which gives the layup angle as a function of the through-the-thickness coordinate is the set of step functions. 
\section{Problem statement and inaccuracy of the proof in \cite{Grenestedt}}
\setcounter{equation}{0}
The lamination parameters are given by the formulas: 
\begin{equation}
\label{eq1}
\xi^A_{[1,2,3,4]}= \frac{1}{2} \int_{-1}^1 [\cos 2 \theta(z), \cos 4 \theta(z), \sin 2 \theta(z), \sin 4 \theta(z)] dz,
\end{equation}
\begin{equation}
\label{eq2}
\xi^B_{[1,2,3,4]}= \int_{-1}^1 [\cos 2 \theta(z), \cos 4 \theta(z), \sin 2 \theta(z), \sin 4 \theta(z)] z dz,
\end{equation}
\begin{equation}
\label{eq3}
\xi^D_{[1,2,3,4]}= \frac{3}{2} \int_{-1}^1 [\cos 2 \theta(z), \cos 4 \theta(z), \sin 2 \theta(z), \sin 4 \theta(z)] z^2 dz,
\end{equation}
\cite{Grenestedt}, equation (13),  where $z$ is the normalized through-the-thickness coordinate and $\theta$: $z \in [-1,1] \mapsto \theta(z) \in \RR$ (where $\RR$ is the set of real numbers) is the (measurable) function which gives the layup angle as a function of $z$. 
If $\theta$ is a measurable function  from $[-1,1]$ into $\RR$, let us call $\xi[\theta]$ the family of associated lamination parameters.

In order to prove the convexity of the set of lamination parameters, one must show that if $\theta_1$ and $\theta_2$ are two measurable functions belonging to some class of functions from $[-1,1]$ into $\RR$ and if $\alpha \in [0,1]$ (one can replace "$\alpha \in [0,1]$" by  "$\alpha \in (0,1)$") then there exists $\theta$ belonging to the same class of functions such that
\begin{equation}
\label{eq9}
(1- \alpha) \xi[\theta_1] + \alpha \xi[\theta_2] = \xi[\theta].
\end{equation}

In \cite{Grenestedt}, p.317,  given $\alpha \in [0,1]$ and two functions $\theta_1$ and $\theta_2$ from $[-1,1]$ into $\RR$, whose regularity is not specified, a sequence of functions $(\theta^n)$ is constructed such that $\xi[\theta^n]$ $\rightarrow$ $(1- \alpha) \xi[\theta_1] + \alpha \xi[\theta_2]$ when $n \rightarrow + \infty$. 
But by no means this proves the convexity of the set of lamination parameters because this does not prove the existence of a function $\theta$ such that \eqref{eq9} is satisfied.
If the sequence of functions $(\theta^n)$ had a uniform limit $\theta$ on $[-1,1]$, it could be said that $\xi[\theta^n] \rightarrow \xi[\theta]$ when $n \rightarrow +\infty$ and the convexity would be proved in case where the functions $\theta^n$ and $\theta$ are in the same class of functions. But as will be proved in the sequel, the sequence $\theta^n$ has not even a simple limit on the set of $x \in [-1,1]$ such that $\theta_1(x) \neq \theta_2(x)$.

Assume that $\theta_1$ and $\theta_2$ are two measurable functions from $[-1,1]$ into $\RR$ and $\alpha \in (0,1)$.
Let us recall the construction of the sequence $(\theta^n)$ in \cite{Grenestedt}. If $n$ is an integer $\geq 1$ (that is $n \in \NN^*$, where $\NN$ is the set of natural numbers, and $\NN^* = \NN \setminus \{0\}$), consider the sequence of $n+1$ points $x_i^n = -1+ 2i/n$, $i=0, \ldots, n$ of the interval $[-1,1]$ and the sequence of $n$ intervals (of the same length $2/n$) $J_i = (x_i^n, x_{i+1}^n)$, $i=0, \ldots, n-1$. For all $n \in \NN^*$, on each interval $J_i$, $\theta^n$ is defined by $\theta^n(x)= \theta_1(x)$ if $x \in (x_i^n, x_i^n + 2\alpha/n)$ and $\theta^n(x)= \theta_2(x)$ if $x \in (x_i^n + 2\alpha/n, x_{i+1}^n)$,  $i=0, \ldots,n-1$. The function $\theta^n$ is not defined for other points of the interval $[-1,1]$ (in finite number), that is the points $x_i^n$ ($i= 0, \ldots, n$) and $x_i^n + 2 \alpha/n$ ($i=0,\ldots,n-1$). 

If $x \in \RR$, we shall denote by $[x]$ the integer part of $x$, that is the unique integer $n \in \ZZ$ (where $\ZZ$ is the set of integers) such that $n \leq x<n+1$ and by $(x)$ the fractional part of $x$ defined by $(x)= x - [x]$.

If $x \in (-1,1)$, for all integer $n \geq 1$, there exists a unique integer $i$, $0 \leq i \leq n-1$, such that 
\begin{equation}
\label{eq9a}
x_i^n \leq x < x_{i+1}^n.
\end{equation}
If we set $y = (x+1)/2$, these relations are equivalent to
\begin{equation}
\label{eq9b}
0 \leq ny -i <1
\end{equation}
or $i = [ny]$.
Similarly for $i=0, \ldots, n-1$, the conditions  
\begin{equation}
\label{eq9c}
x_i^n  < x < x_i^n + 2 \alpha/n
\end{equation}
and 
\begin{equation}
\label{eq9d}
x_i^n + 2 \alpha/n < x <  x_{i+1}^n
\end{equation}
are equivalent to 
\begin{equation}
\label{eq9e}
0 < ny -i  <\alpha\, (\Leftrightarrow 0 < (ny) < \alpha)
\end{equation}
and
\begin{equation}
\label{eq9f}
\alpha < ny -i  <1\, (\Leftrightarrow\alpha < (ny) <1).
\end{equation}
Let us recall Kronecker's Theorem: if $\theta$ is an irrational number (that is $\theta \in \RR \setminus \QQ$, where $\QQ$ is the set of rational numbers), the set of fractional values $(n \theta)$ = $n\theta - [n \theta]$ of $n\theta$, $n \in \NN$, is dense in $[0,1]$ (\cite{Hardy-Wright}, p.502, Theorem 439). This means that for all $z \in [0,1]$, for all $\epsilon >0$, there exists $n \in \NN$ such that $|z - (n\theta)| < \epsilon$.
We begin with the following lemma.

\begin{lemme}
\label{lemma1}
Let $p,q \in \NN$ such that $p$ and $q$ are relatively prime and  $0 <p <q$. Then there exists an infinity of $n$ and $i$  $\in \NN$, such that $n \geq q-1 \geq 1$, $0 \leq i \leq n-1$ and $np - qi =1$.
Moreover for all $j \in \NN$, $1 \leq j \leq q-1$, there exists an infinity of $n'$, $i'$ $\in \NN$ such that $n' \geq j(q-1) \geq 1$, $0 \leq i' \leq n'-1$ and $n'p - qi' =j$.
\end{lemme}
We have $p\geq 1$, $q \geq 2$, $p \leq q-1$.
Since $p$ and $q$ are relatively prime, Bezout's Theorem  shows that there exists $n_0$ and  $i_0$  $\in \ZZ$ such that $n_0p - qi_0 =1$. If $n_0$, $i_0$ $\in$ $\ZZ$ is a particular solution of the equation $np - qi =1$, the other solutions of this equation are under the form $n= n_0 + k q$, $i= i_0 +kp$, $k \in \ZZ$.
Then there exists an infinity of $n$ and $i$  $\in \NN$, such that $n \geq q-1$ and $np - qi =1$. If $np - qi =1$ then 
$i = (np- 1)/q$ and  $i \geq (n-1)/q \geq 0$, $i \leq (n(q-1) -1)/q$ $\leq n - (n+1)/q$  $\leq$ $n-1$. 
If $j \in \NN$, $1 \leq j \leq q-1$, multiply the relation $np - qi =1$ by $j$  and set $n' = nj$, $i'= ij$. We get $n' \geq j(q-1)$, $0 \leq i' \leq j(n-1)$ $\leq n' -1$ and $n' p - q i' = j$.
$\Box$

The following lemma shows that for all $x \in [-1,1]$ such that $\theta_1(x) \neq \theta_2(x)$, the sequence $\theta^n(x)$ has no limit when $n \rightarrow + \infty$. 
\begin{lemme}
\label{lemma2}
Let $x \in [-1,1]$ and set $y=(x+1)/2$ $\in$ $[0,1]$.  Let $\alpha \in (0,1)$.

If $x \in \RR\setminus \QQ$, there exists an infinity of $n \in \NN$ such that  $0 < (n y) < \alpha$ and there exists an infinity of $n \in \NN$ such that  $\alpha < (n y) < 1$.
If $\theta_1(x) \neq \theta_2(x)$, $\theta^n(x)$ has no limit when $n \rightarrow + \infty$.

If $x$ $\in \QQ$, there exists an infinity of $n \in \NN$ such that $\theta^n(x)$ is not defined.

Moreover, if $x$ $\in \QQ \cap (-1,1)$, then $y=(x+1)/2 \in \QQ\, \cap\, (0,1)$ can be written under the form $y =p/q$ where $p\in \NN$, $q \in \NN^*$, $p$ and $q$ are relatively prime and  $0 <p <q$. If $q$ is such that $1/q < \alpha$ and $1/q < 1-\alpha$, there exists an infinity of $n \in \NN$ such that  $0 < (n y) < \alpha$ and there exists an infinity of $n \in \NN$ such that  $\alpha < (n y) < 1$.
If $\theta_1(x) \neq \theta_2(x)$, $\theta^n(x)$ has no limit when $n \rightarrow + \infty$.

It follows that for all $x \in [-1,1]$ such that $\theta_1(x) \neq \theta_2(x)$,  $\theta^n(x)$ has no limit when $n \rightarrow + \infty$.
\end{lemme}
{\bf Proof}
The case $x \in \RR \setminus \QQ$ follows from Kronecker's Theorem.

If $x$ $\in \QQ$, then  $x$ can be written under the form $x =p/q$ where $p \in \ZZ$, $q \in \NN^*$, $p$ and $q$ are relatively prime and  $-q \leq p \leq q$. If we set $n = k(2q)$ and $i =k(p+q)$, $k \in \NN^*$, then $n \geq 1$, $0 \leq i \leq n$ and $x = x_i^n$, so that $\theta^n(x)$ is not defined.  

Let $x \in \QQ \cap (-1,1)$. Then $y=(x+1)/2 \in \QQ\, \cap\, (0,1)$ can be written under the form $y =p/q$ where $p, q \in \NN$, $p$ and $q$ are relatively prime and  $0 <p <q$.
From Lemma \ref{lemma1} for all $j  \in \NN$ such that $1 \leq j \leq q-1$, there exists an infinity of $n \in \NN$, $n \geq 1$ and $i \in \NN$, $0 \leq i \leq n-1$ such that $n y - i = j/q$. In other words  for all $j  \in \NN$ such that $1 \leq j \leq q-1$, there exists an infinity of $n \in \NN$, $n \geq 1$ such that $(ny) =j/q$. Now assume that $q$ is such that $1/q < \alpha$ and $1/q < 1-\alpha$. Then there exists an infinity of $n \in \NN$, $n\geq 1$ such that $(ny) =1/q$ and an infinity of $n \in \NN$, $n\geq 1$ such that $(ny) =1-1/q$. Since $0 < 1/q < \alpha$ and $\alpha < 1-1/q$, it follows that there exists an infinity of $n \in \NN$ such that  $0 < (n y) < \alpha$ and there exists an infinity of $n \in \NN$ such that  $\alpha < (n y) < 1$.
$\Box$

\section{Proof of the convexity of the set of lamination parameters}

We shall prove the convexity of the set of lamination parameters when the functions $\theta$ are in the class of step functions.
Recall that a function $\theta$: $z \in [-1,1] \mapsto \theta(z) \in \RR$ is a step function if there exists a sequence $(a_i)_{i=0,\ldots,N}$, $a_0 = -1$, $a_N=1$, $a_i < a_{i+1}$, $i=0,\ldots,N-1$ such that $\theta$ is constant in each of the intervals $(a_i, a_{i+1})$, $i=0,\ldots,N-1$. 

Let $\theta_1$ and $\theta_2$ be two step functions on $[-1,1]$.
One can find a sequence $(a_i)_{i=0,\ldots,N}$, $a_0 = -1$, $a_N=1$, $a_i < a_{i+1}$, $i=0,\ldots,N-1$ such that the two step functions $\theta_1$ and $\theta_2$ are constant in each of the intervals $(a_i, a_{i+1})$, taking the values $\theta_1^i$ and $\theta_2^i$, $i=0,\ldots,N-1$. 

The lamination parameters are all under the form 
\begin{equation}
\label{eq10}
\int_{-1}^1 f(\theta(z)) z^j  dz,\, j=0,1,2
\end{equation}
where $f$ is a continuous function on $\RR$.
We have
\begin{equation}
\label{eq11}
\int_{-1}^1 f(\theta_k(z)) z^j dz = \sum_{i=0}^{N-1}f(\theta_k^i)\int_{a_i}^{a_{i+1}} z^j dz,\, k=1,2, \, j=0,1,2.
\end{equation}
Then if $\alpha \in [0,1]$, $j=0,1,2$,
\begin{equation}
\label{eq12}
(1-\alpha)\int_{-1}^1 f(\theta_1(z)) z^j dz + \alpha \int_{-1}^1 f(\theta_2(z)) z^j dz= \sum_{i=0}^{N-1} [(1- \alpha)f(\theta_1^i) + \alpha f(\theta_2^i)] \int_{a_i}^{a_{i+1}}  z^j dz.
\end{equation}
In order to prove the convexity of the set of lamination parameters, one must find a step function $\theta$ on $[-1,1]$ such that
\begin{equation}
\label{eq13}
\sum_{i=0}^{N-1} [(1- \alpha)f(\theta_1^i) + \alpha f(\theta_2^i)] \int_{a_i}^{a_{i+1}}  z^j dz = \int_{-1}^{1}  f(\theta(z))z^j dz, \, j=0,1,2. 
\end{equation}
Let us search $\theta$ such that in each interval $(a_i,a_{i+1})$, $\theta$ takes the values $\theta^i_1$ (resp. $\theta^i_2$) on a (disjoint) union of open intervals $E^i_1$ (resp. $E^i_2$): $\forall i=0,\ldots, N-1,$
\begin{equation}
\label{eq14}
\theta(z)= \theta^{i}_1 1_{E^i_1}(z)+  \theta^i_2 1_{E^i_2}(z), \, z \in (a_i, a_{i+1}),
\end{equation} 
\begin{equation}
\label{eq15}
\overline{E}^i_1  \cup \overline{E}^i_2 = [a_i,a_{i+1}],  \,  E^i_1 \cap E^i_2 = \emptyset. 
\end{equation} 
In \eqref{eq14}, we have used the following notation: if $E$ is a subset of $\RR$, $1_E$ is the indicator function of the subset $E$, that is the function such that $1_E(x)=1$ if $x \in E$ and $1_E(x)= 0$ if $x \not \in E$. Equations \eqref{eq14} and \eqref{eq15} imply
\begin{equation}
\label{eq16}
\int_{a_i}^{a_{i+1}}  f(\theta(z))z^j dz = f(\theta^i_1) \int_{E^i_1} z^j dz + f(\theta^i_2) \int_{E^i_2} z^j dz,\, i=0,\ldots, N-1,\, j=0,1,2. 
\end{equation}
It is sufficient to find $E^i_1$ (and then $E^i_2$) such that
\begin{equation}
\label{eq17}
[(1- \alpha)f(\theta_1^i) + \alpha f(\theta_2^i)] \int_{a_i}^{a_{i+1}}  z^j dz = f(\theta^i_1) \int_{E^i_1} z^j dz + f(\theta^i_2) \int_{E^i_2} z^j dz,  \, j=0,1,2. 
\end{equation}
It is enough to verify
\begin{equation}
\label{eq18}
(1- \alpha) \int_{a_i}^{a_{i+1}}  z^j dz =  \int_{E^i_1} z^j dz,  \, i=0,\ldots, N-1,\, j=0,1,2 
\end{equation}
and
\begin{equation}
\label{eq19}
\alpha \int_{a_i}^{a_{i+1}}  z^j dz =  \int_{E^i_2} z^j dz,  \, i=0,\ldots, N-1,\, j=0,1,2. 
\end{equation}
But if ${E^i_1}$ and  ${E^i_1}$ verify \eqref{eq15} then for all $i=0,\ldots, N-1, \, j=0,1,2,$
\  
\begin{equation}
\label{eq20}
\int_{E^i_1} z^j dz+ \int_{E^i_2} z^j dz = \int_{a_i}^{a_{i+1}}  z^j dz =  (1- \alpha) \int_{a_i}^{a_{i+1}}  z^j dz + \alpha \int_{a_i}^{a_{i+1}}  z^j dz,
\end{equation}
thus it suffices to find $E^i_1$ verifying \eqref{eq18} and \eqref{eq19} will be necessarily verified.  The answer to this problem is given by the following lemma:
\begin{lemme}
\label{lemma3}
Let $A$, $B$ $\in \RR$, $A < B$ and  let $\alpha \in \RR$, $0 < \alpha < 1$. Then there exists an union of intervals $\subset (A,B)$ (denoted $E$) such that
\begin{equation}
\label{eq21}
\alpha \int_A^B z^j dz = \int_{E} z^j dz, \, j=0,1,2.
\end{equation}
\end{lemme}
{\bf Proof}
Since $E$ must meet the three conditions \eqref{eq21}, we seek $E$ under the form $E = (a,b) \cup (c,d)$ with $a,b,c,d$ $\in \RR$, $A<a<b<c<d<B$ and from the begining we enforce the condition $b-a$ = $d-c$ (denoted $e$), so that $E$ will depend on three parameters to be determined. One must verify the conditions: 
\begin{equation}
\label{eq22}
\alpha \int_A^B z^j dz = \int_{a}^b z^j dz + \int_c^d z^j dz, \, j=0,1,2,
\end{equation}
that is
\begin{equation}
\label{eq23}
\alpha (B-A)=  b-a + d-c  = 2 e,
\end{equation}
\begin{equation}
\label{eq24}
\alpha (B^2-A^2)=  b^2-a^2 + d^2-c^2,
\end{equation}
\begin{equation}
\label{eq25}
\alpha (B^3-A^3)=  b^3-a^3 + d^3-c^3.
\end{equation}
Equation \eqref{eq23} makes it possible to determine $e$: $e= \alpha(B-A)/2$. Setting $X= (a+b)/2$ and  $Y= (c+d)/2$, we get
\begin{equation}
\label{eq26}
a= X -e/2,\, b = X+ e/2, \, c= Y -e/2, d= Y+e/2,
\end{equation}
\begin{equation}
\label{eq27}
\alpha (B^2-A^2)=  \alpha(B-A)(B+A) = 2e(A+B),
\end{equation}
\begin{equation}
\label{eq28}
b^2-a^2 + d^2-c^2 = (b-a)(a+b)+(d-c)(c+d) = 2e(X+Y).
\end{equation}
Equations  \eqref{eq24}, \eqref{eq27}, \eqref{eq28} give 
\begin{equation}
\label{eq29}
X+Y= A+B.
\end{equation}
On the other hand
\begin{equation}
\label{eq30}
b^3-a^3 = (X+e/2)^3 - (X-e/2)^3 = 2[3 X^2 (e/2) + (e/2)^3]= e (3 X^2 + e^2/4),
\end{equation}
\begin{equation}
\label{eq30a}
d^3-c^3 = (Y+e/2)^3 - (Y-e/2)^3 = 2[3 Y^2 (e/2) + (e/2)^3]= e (3 Y^2 + e^2/4),
\end{equation}
\begin{equation}
\label{eq31}
A= (A+B)/2 - (B-A)/2 = (X+Y)/2 -e/{\alpha},
\end{equation}
\begin{equation}
\label{eq32}
B= (A+B)/2 + (B-A)/2 = (X+Y)/2 +e/{\alpha},
\end{equation}
\begin{eqnarray}
\label{eq33}
&&B^3-A^3 = [(X+Y)/2+e/\alpha]^3 - [(X+Y)/2-e/\alpha]^3 = \nonumber \\
&& 2[3 ((X+Y)/2)^2 (e/\alpha) + (e/\alpha)^3]= 2e/\alpha [3/4 (X+Y)^2 + (e/\alpha)^2].
\end{eqnarray}
Then equation \eqref{eq25} can be written
\begin{eqnarray}
\label{eq34}
2e [3/4 (X+Y)^2 + (e/\alpha)^2] = e (3 X^2 + e^2/4) + e (3 Y^2 + e^2/4)
\end{eqnarray}
that is
\begin{eqnarray}
\label{eq35}
2 [3/4 (X+Y)^2 + (e/\alpha)^2] = 3 (X^2 + Y^2) + e^2/2
\end{eqnarray}
or
\begin{equation}
\label{eq36}
3[X^2 + Y^2 -(X+Y)^2/2] =  e^2 (2/\alpha^2 -1/2)
\end{equation}
and
\begin{equation}
\label{eq37}
(X-Y)^2 = (e^2/3)(4/\alpha^2 -1).
\end{equation}
Since  $4/\alpha^2 -1 >0$, let us set 
\begin{equation}
\label{eq38}
\beta = (e/\sqrt{3}) \sqrt{4/\alpha^2 -1} >0.
\end{equation}
Since $Y> X$, from \eqref{eq29} and \eqref{eq37} we get
\begin{equation}
\label{eq38a}
Y-X = \beta,
\end{equation}
\begin{equation}
\label{eq39}
X = [A+B - \beta]/2,
\end{equation}
\begin{equation}
\label{eq40}
Y = [A+B + \beta]/2.
\end{equation}
One must now verify that the solution $a,b,c,d$ we have obtained (see formulas \eqref{eq26}) is suitable, that is  $A <a<b<c<d<B$. This amounts to show that $Y-X >e$ and  $Y-X + e < B-A$.
Since $0<\alpha <1$, then $Y-X$ = $\beta >e$. On the other hand the relation $Y-X + e < B-A$ is equivalent to
\begin{equation}
\label{eq41}
(e/\sqrt{3}) \sqrt{4/\alpha^2 -1} +e < 2 e/\alpha
\end{equation}
that is
\begin{equation}
\label{eq42}
\sqrt{4/\alpha^2 -1} < \sqrt{3} (2 /\alpha -1)
\end{equation}
or
\begin{equation}
\label{eq43}
{4/\alpha^2 -1} < 3(2 /\alpha -1)^2.
\end{equation}
Equation \eqref{eq43} is equivalent to the condition
\begin{equation}
\label{eq44}
\alpha^2 -3 \alpha +2 > 0
\end{equation}
which is verified since if $0 < \alpha <1$ then $\alpha^2 -3 \alpha +2 = (1-\alpha) (2-\alpha) >0$.
${\Box}$ 
%
%

%
\end{document}